\documentclass[aps,prl,groupedaddress,showpacs,twocolumn,floatfix]{revtex4}
\usepackage{amsmath}
\usepackage{amssymb}
\usepackage{graphicx}
\usepackage{epsfig}

\begin{document}


\title{Two-dimensional Bose gas under extreme rotation}
\author{S. Sinha$^{1}$ and G. V. Shlyapnikov$^{2,3}$}
\address{\mbox{$^{1}$ Max-Planck-Institut f\"{u}r Physik Komplexer 
Systeme, 38 N\"{o}thnitzer Stra\ss e, 01187-Dresden, Germany}\\
\mbox{$^{2}$Laboratoire Physique Th\'eorique et Mod\`eles Statistiques,
B\^at.100, Universit\'e Paris-Sud, 91405 Orsay, France}\\
\mbox{$^{3}$Van der Waals - Zeeman Institute, University of Amsterdam, 
Valckenierstraat 65/67, 1018 XE Amsterdam, The Netherlands}
}

 
\begin{abstract}
We show that a Bose-condensed gas under extreme rotation in a 2D 
anisotropic trap, forms a novel elongated quantum fluid which has a 
roton-maxon excitation 
spectrum. For a sufficiently large interaction strength, the roton energy 
reaches zero and the system undergoes a second order quantum transition to 
the state with a periodic structure - rows of
vortices. The number of rows increases with the 
interaction, and the vortices eventually form a triangular Abrikosov lattice.

\end{abstract}

\maketitle
\narrowtext

Rotating Bose Einstein condensates (BEC) of trapped atoms constitute a
novel many-body system where nucleated quantized vortices form a triangular 
lattice \cite{rot1}, and a fast rotation is expected to
change dramatically the properties of the gas.
In a harmonically trapped BEC rotating at
a frequency $\Omega$ close to the trap frequency in the rotation plane 
$\omega_{\perp}$, the vortex lattice can melt when the number of vortices 
approaches the number of particles \cite{fetter,cooper}.
In this respect, there is an analogy with type-II superconductors, which 
undergo a transition to the normal phase above a critical value of the 
magnetic field \cite{tinkham}.
Possible signatures of melting of the vortex lattice in rapidly rotating BEC's
were observed in experiments at JILA and at ENS \cite{ENS2}.
The proposals that are put forward to describe the state
of the rapidly rotating Bose gas include yrast states\cite{mottelson}, 
correlated quantum Hall states
of bosons \cite{cooper,hall}, and a giant vortex state \cite{baym}.

In this letter we show that an anisotropy in the trapping potential can drastically
change the picture. As found at ENS \cite{rosen}, a 
Bose gas under critical rotation ($\Omega \sim \omega_{\perp}$) can 
become very elongated in one direction in the rotating 
frame. We focus on this case and assume that $\Omega$ is
close to the smallest of the trap frequencies in the rotation plane. Then the
condensate becomes free along the direction of the weaker confinement 
and forms a novel quantum fluid in a narrow channel. The excitation
spectrum of this fluid has a ''roton-maxon'' character and
becomes unstable at a critical interaction
strength. This instability leads to the
formation of a periodic structure which represents vortex rows. 
An increase in the interaction (decrease in the anisotropy) increases
the number of rows and reduces the correlation length, and the
gas ultimately enters the strongly correlated regime. 

We consider a two-dimensional (2D) Bose-condensed gas at zero 
temperature, 
rotating with frequency $\Omega$ and harmonically trapped with frequencies $\omega_{x,y} =
\omega_{\perp}\sqrt{1\mp\epsilon}$ along the $x,y$ axes in the rotating frame. The Hamiltonian 
of the system in this frame reads (see \cite{stringari}): 
\begin{eqnarray}      \label{Hamil1}
H\!=\!\!\!\int\!\! d^{2}r\tilde{\Psi}^{\dagger}\!\left[\!\frac{(-i\hbar
\vec{\nabla}\!+\!m \vec{r}\!\times\!\vec{\Omega})^2}{2m}\!+\! 
V_{eff}(\vec{r})\!+\!\frac{g}{2}\tilde{\Psi}^{\dagger}\tilde{\Psi}\right]\!\tilde{\Psi},
\end{eqnarray}
where $\tilde\Psi^{\dagger}$,$\tilde\Psi$ are bosonic field operators, $m$ 
is the particle mass,
$g$ is the coupling constant for the mean-field interaction \cite{g}, 
and the effective trapping potential is:
\begin{equation}    \label{Veff}
V_{eff}(\vec{r}) =(m/2)[(\omega_x^2-\Omega^2)x^2+(\omega_y^2-\Omega^2)
y^2].
\end{equation}
The Hamiltonian (\ref{Hamil1}) is analogous to that of charged particles
in the magnetic field, and in this respect the quantity $m \vec{r}\times
\vec{\Omega}$ is the gauge field. We consider the limit of extreme rotation, 
where $\Omega =\omega_x\equiv\omega_{\perp}\sqrt{1 - \epsilon}$ and the gas 
becomes free along the $x$ direction. Then, assuming that in this direction 
the atoms are confined in a large rectangular box of size $L$, the gas 
becomes a long cigar. After the gauge 
transformation $\tilde{\Psi} = \Psi e^{im\Omega xy/\hbar}$, the 
Hamiltonian in
the Landau gauge can be written as:
\begin{eqnarray}        \label{Hamil2}
H\!=\!\!\!\int\!\! d^{2}r \Psi^{\dagger}\!\left[\!\frac{(p_x\!+\!2m\Omega y)^2\!+\!p_y^2}{2m}\!+\!
\frac{m\omega_{-}^{2} y^2}{2}\!+\!\frac{g}{2}\Psi^{\dagger}\Psi\right]\!\Psi, 
\end{eqnarray}
where $\omega_{-} = \omega_{\perp}\sqrt{2 \epsilon}\ll\Omega$, assuming a 
small ellipticity $\epsilon$.

Omitting the interaction term, the single particle eigenstates are 
the Landau levels \cite{LL} separated from each
other by an
energy gap $\sim 2 \hbar \Omega$. In the dilute limit,  where the mean
field interaction $g n_{2D} \ll \hbar \Omega$, ($n_{2D}$ is the 2D
density), we may restrict our discussion within the lowest Landau level. 
Then the field operator can be written in the form $\Psi = \sum_{k}
\phi_{k} a_{k}$, where $a_{k}$
is the creation operator of a particle with momentum $k$ along the x
direction and $\phi_{k}$ is the corresponding eigenfunction:
\begin{equation} \label{eigenf}
\phi_{k}(x,y)  =  \frac{\exp(ikx)}{(\pi l_0^2L^2)^{1/4}}
\exp\left\{-\frac{1}{2}\left(\frac{y}{l_0}+\frac{\Omega
kl_0}{\tilde\Omega}\right)^2\right\},
\end{equation}
where $\tilde{\Omega} = \sqrt{\Omega^{2} + \omega_{-}^{2}/4}$, and $l_{0}
= (\hbar/2m\tilde{\Omega})^{1/2}$.
Then, after the spatial integration of Eq.~(\ref{Hamil2}), we obtain an
effective one-dimensional (1D) Hamiltonian:
\begin{eqnarray} \label{effective}
& & H =  
\sum_{k}(\hbar^{2}k^{2}/2 m^*) a^{\dagger}_{k}a_{k}+(g^*/2 L)
\sum_{k,k',q} a^{\dagger}_{k+q}a^{\dagger}_{k'-q}a_k a_{k'}  \nonumber   \\ 
& &\exp\left\{-l_0^2[(k - k' + q)^{2}
+ q^{2}]/2\right\},
\end{eqnarray} 
where $g^* = g/\sqrt{2 \pi}l_0$ is an effective 1D coupling constant.
The Hamiltonian (\ref{effective}) describes particles with a large 
effective 
mass $m^* = m(2\tilde{\Omega}/\omega_-)^2\gg m$. The fact that $m^*$
is not infinite and the kinetic energy term is still present originates from the asymmetry of
the trapping potential. This asymmetry leads to a small difference
between the frequencies $\Omega$ and $\tilde\Omega$. However, once the finite kinetic
energy term is extracted one may put $\tilde\Omega=\Omega$, and we have done this in
the second term on the rhs of Eq.(\ref{effective}). 

The momentum dependence of the interaction term in the Hamiltonian (\ref{effective})
originates from the presence of the gauge field. The wavefunctions of particles which
have opposite momenta in the $x$ direction are shifted in opposite directions along the
$y$ axis, which decreases their overlap and reduces the interaction amplitude.  
 
The behavior of the system is governed by the particles with momenta
$k\alt
l_{0}^{-1}$, for which the extension of the wave function in the y
direction is $l_0$. If the 1D density $n
 = N/L$ (N is the total number
of particles) satisfies the condition $n l_0 \ll 1$, then we are dealing
with a 1D Bose gas. In this case, characteristic momenta that are
important are at least of the order of $n^{-1}$ or smaller. They
satisfy the condition $kl_0 \ll 1$, and the exponential term
in Eq.(\ref{effective}) is equal to unity. Then the
Hamiltonian(\ref{effective}) corresponds to the Lieb-Liniger model for the
1D Bose gas, well described in literature\cite{lieb}.

We, therefore, focus on the other extreme, where
\begin{equation} \label{condition1}
n l_0 \gg1. 
\end{equation}
Then the system can be viewed as a 2D gas in a narrow channel. The 2D
density is $\sim n/l_0$ and the interaction energy per particle is
$I\sim ng/l_0$. The characteristic kinetic energy of a
particle at the mean distance from other particles is $K \sim \hbar^2
n/m^*l_0$, and for $K\gg
I$ the wave function of the particle at such interparticle distances is 
not influenced by the interactions. The gas is then in the weakly
interacting mean-field regime. The criterion of weak interactions 
takes the form (see, for example \cite{petrov}):
\begin{equation} \label{condition2}
m^* g/\hbar^2 \ll 1.
\end{equation}
The correlation length is $l_c 
\sim
\hbar/\sqrt{m^* g^* n}$, and for $l_c \gg l_0$ the
gas enters the 1D regime. In this case the 1D criterion of weak
interactions, $m^* g^*/\hbar^2 n \ll 1$, is automatically
satisfied under the conditions (\ref{condition1}) and (\ref{condition2}).

For the weakly interacting 2D Bose gas in a narrow channel, as well as in
the 1D Bose gas, for a sufficiently large size $L$ the ground state can
be a quasi-condensate. In this state the density fluctuations are suppressed, 
but the phase fluctuates in the $x$ 
direction on a distance scale greatly exceeding 
$l_c$ \cite{petrov}. However, locally the quasi-condensate is
indistinguishable from a true BEC. As follows 
from the analysis of hydrodynamic equations in the density-phase 
representation \cite{petrov}, the excitation spectrum is the same as the 
one 
obtained in the Bogoliubov approach assuming that most particles are in the
condensate. Employing this approach we first reduce the Hamiltonian
(\ref{effective}) to a bilinear form:
\begin{eqnarray} \label{bog1}
H_b & = &  
\sum_{k}\left[\hbar^2 k^2/2 m^*
+ 2ng^* \exp{\{-k^2l_{0}^{2}/2\}} \right]
a^{\dagger}_{k} a_{k} \nonumber \\ 
& + &
(ng^*/2) 
\sum_{k}\exp\left\{-k^2l_{0}^{2}\right\}( a^{\dagger}_{k}
a^{\dagger}_{-k} + a_{k} a_{-k})
\end{eqnarray}
Then, diagonalizing the Hamiltonian (\ref{bog1}) we obtain the excitation spectrum:
\begin{eqnarray} \label{bog2}
\epsilon^{2}(k) & = & \left[\hbar^2 k^2/2m^* + 
g^* n \left\{ 2
\exp\left(-k^2l_{0}^2/2\right) 
- 1\right\}\right]^{2} \nonumber \\
& & -
g^{*2}n^{2}
\exp\left\{-2k^2l_{0}^2\right\}.
\end{eqnarray}

The key feature of the excitation energy (\ref{bog2}) 
is the momentum dependence of the interaction terms proportional to $g^*$.  
The structure of the spectrum depends on the ratio of the mean-field
interaction to the kinetic energy at momentum $k\approx 1/l_0$. 
This ratio can take both small and large values and is given by:
\begin{equation}      \label{beta}
\beta = \frac{n g^*}{\hbar^2/2 m^*l_0^2} = \sqrt{\frac{2}{\pi}}\,
\left(\frac{m^*g}{\hbar^2}\right) nl_0
\end{equation}

In units of $m\hbar \tilde{\Omega}/m^*$, the excitation energy is a universal function
of $\beta$ and $kl_0$. For small $\beta$, the interaction terms proportional to $g^*$
in Eq.(\ref{bog2}) are important only at $k\ll 1/l_0$, where they become 
momentum independent. Then Eq.(\ref{bog2}) gives the ordinary Bogoliubov spectrum,
with a small sound velocity $c_s=\sqrt{ng^*/m^*}$.
\begin{figure}[htb]
\begin{center}
\epsfig{file=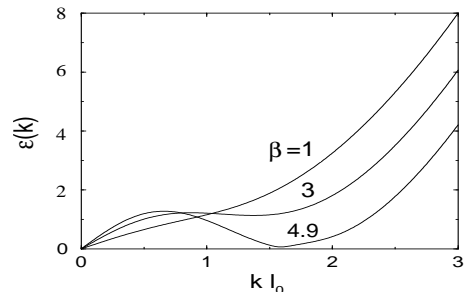,
height=4cm,width=6cm}
\end{center}
\hskip -2cm
\vspace*{-1.2cm}
\caption{Excitation energy (in units of $m\hbar \tilde{\Omega}/m^*$) 
versus $k l_0$.}
\label{fig:1}  
\end{figure}

For $\beta\gtrsim 1$ the situation drastically changes. The interaction is already important for momenta
$k\gtrsim 1/l_0$, where the interaction dependent terms in Eq.(\ref{bog2})
decrease with increasing k. For this reason the spectrum develops the
roton-maxon structure for $\beta > 2.6$ (see Fig.1). 
This structure is well-known in the physics of superfluid $^{4}$He and is 
predicted for trapped dipolar condensates \cite{gv} and studied for lattice bosons
with long-range interactions \cite{kov}. 
The roton minimum is located at $k\sim 1/l_0$
and the corresponding excitation energy decreases with increasing
$\beta$.
For a critical value $\beta = 4.9$, the roton minimum reaches zero at
$k = k_c = 1.6/l_0$ and a further increase in $\beta$ makes the
system unstable. We thus conclude that our weakly interacting 2D
Bose-condensed gas in a narrow channel is stable for $\beta \le 4.9$, 
exhibiting a roton-maxon spectrum in the range 
$2.6<\beta<4.9$.
For $\beta>4.9$, where the Bose-condensed state is unstable, one has 
to find a new ground state. 

At the instability point the excitations with momenta $\pm k_c$ can be
excited without any cost of energy.
This indicates that the ground state macroscopic wavefunction 
can contain several momentum components. A general form of this type
of wavefunction reads: 
\begin{equation} \label{multicomp}
\psi = \sqrt{N}\left[C_0 \phi_0 + \sum_{i=1}^{j}(C_{k_i}\phi_{k_i}
+ C_{-k_i}\phi_{-k_i})\right].
\end{equation}
The absence of current in the $x$-direction requires $|C_{k_i}| =
|C_{-k_i}|$, and the normalization condition reads $|C_0|^2 + 2 \sum_{i}^{j} |C_{k_i}|^2 = 1$.
Note that Eq.(\ref{multicomp}) gives two possibilities: a
$(2j+1)$ component wavefunction with $C_0 \ne 0$, and a $2j$ component
wavefunction for which $C_0 = 0$.

To understand the instability of the single component state, we 
consider a wavefunction with three components:
\begin{equation}  \label{3c}
\psi = \sqrt{N}[C_{k} \phi_{k} + C_{0} e^{i \theta} \phi_{0} + C_{-k}
\phi_{-k}].
\end{equation}
Here all the $C$-coefficients are real, and we may put $C_k=C_{-k}$. The quantity
$|C_{k}|^{2}$ represents the occupation number of the mode with
momentum $k$. Normalization of the wavefunction requires $ 2 |C_{k}|^{2} + C_{0}^{2} = 1$.
We then find the critical value $\beta_c$ of the parameter $\beta$,
above which this 3-component state has lower energy than the single
component one. The wavefunction (\ref{3c}) leads to $a_0=C_0$, $a_{\pm k}=C_k$ in the
Hamiltonian(\ref{effective}), and we obtain the energy per particle: 
\begin{equation} \label{en3comp}
E[C_k,k]/N = A(k) |C_{k}|^{4} + B(k) |C_{k}|^{2} + 
g^* n/2,
\end{equation}
where the coefficients $A(k)$ and $B(k)$ are given by:
\begin{eqnarray}
A(k) & = & g^* n[3 - 8 t + 2 t^{4} - 4 t^{2} 
\cos(2\theta)] \label{A} \\
B(k) & = & \hbar^2 k^2/m^* -g^* n[2 - 4 t  
- 2 t^{2}\cos(2\theta)] \label{B},
\end{eqnarray}
with $t = \exp\{-k^2l_0^2/2\}$.
For $|C_{k}|^{2}<1/2$, the energy is 
minimized for $\cos(2\theta) = -1$,  and we have $A(k) >0$.
Minimizing $E$ with respect to $|C_{k}|^{2}$, yields 
$|C_{k}|^{2} =-B(k)/2A(k)$. 
The energy of the 3-component state is 
$$E/N = g^* n/2 - B^2/4 A,$$ 
and it is lower than the energy of the single component state.
The physically acceptable solution requires $B(k)<0$. Therefore,
the transition from the single to 3-component state occurs when the 
minimum value of $B(k)$ reaches zero. Using Eq.(\ref{B}) we find that this
happens at $k = k_c$ and $\beta=\beta_c = 4.9$. So, the 3-component 
solution becomes the ground state for $\beta>4.9$. This breaks the translational symmetry and gives rise to a 
modulation of the density along the $x$-axis with a period of $2 \pi/k_c$, 
which is similar to a scenario proposed for superfluid $^{4}$He flowing along a capillary with a velocity
exceeding the critial Landau velocity \cite{pitaevskii}.

A 3-component wavefunction can be viewed as two vortex rows along the 
$x$-axis. The nodes in the $x,y$ plane are obtained 
straightforwardly, and near the transition point they are very far from the line $y=0$. 
The transition from the single to 3-component wavefunction can be treated 
as a second order quantum transition. The energy and chemical potential $\mu$ 
change continuously, whereas the compressibility 
undergoes a jump at the transition point. For the 3-component state it is smaller by an
amount
\begin{equation}  \label{deltamu}
\Delta\left(\frac{\partial \mu}{\partial 
n}\right) = \frac{2g^*[1-t_c]^4}{[3 - 8t_c + 4 
t_{c}^2 + 2 
t_{c}^4]}\approx 0.5g^*,
\end{equation}
where $t_c = \exp\{-k_c^2l_0^2/2\}\approx 0.28$.

Near the
transition point quantum fluctuations increase due to the vanishing excitation energy at 
the roton minimum. This energy can be expressed as 
\begin{eqnarray}
\epsilon(k) = \left(\frac{\hbar^2}{2m^*l_0^2}\right)\left[\frac{2}{\beta_{c}}(\beta_c -\beta) +  
\gamma l_0^4(k^2 - k_{c}^2)^2\right]^{1/2}\nonumber,
\end{eqnarray}
where $\gamma \approx 0.12$.
For the mean square fluctuations of the density we then obtain:
\begin{eqnarray}  \label{fluctuation}
\left(\frac{\delta n}{n}\right)^2=\int\frac{dk}{2n \pi} 
\left[\frac{\hbar^2k^2}{2m^*\epsilon(k)}-1\right]\approx \frac{1}{nl_{0}} 
\ln\left(\frac{1}{\beta_c-\beta}\right)\nonumber.
\end{eqnarray}
These fluctuations become large for $\delta\beta=(\beta_c-\beta)\alt\exp(-nl_0)$.
Thus, the transition from a single to 3-component wavefunction occurs in the interval
$\delta\beta$, which is exponentially narrow due to the inequality 
(\ref{condition1}). The behavior of the system in this region
requires a special investigation and is beyond the scope of this paper.
\begin{figure}
\begin{center}
\epsfig{file=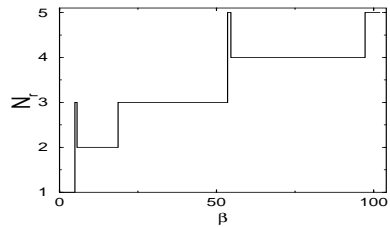,
height=3cm,width=5cm}
\end{center}
\hskip -2cm   
\vspace*{-1.2cm}
\caption{The number of components ($N_r$) in the ground state macroscopic  
wavefunction for various values of $\beta$.
\label{fig:2}}
\end{figure} 

It is important that already for $\beta=5.4$ the ground state wavefunction changes from 
3 to 2-component, and it again becomes 3-component at $\beta=20$. 
With increasing $\beta$ (either increasing $g^* n$ or decreasing 
$\omega_-$), more 
momentum states are macroscopically occupied. This is because 
an increase of $\beta$ is equivalent to increasing the effective 
mass $m^*$, which 
makes momentum states in the lowest Landau level (LLL) more degenerate. 

Our results for the number
of momentum components in the ground state wavefunction are displayed in Fig.2.
These states describe one or 
several vortex 
rows along the x-axis. For example, a two component state
represents one vortex row (see Fig.3 I).
Eventually, a triangular Abrikosov vortex lattice
\cite{abrikosov} is formed when increasing $\beta$ to very large values.
It should be mentioned here that the structure of vortex rows has been obtained in
the studies of type-II superconductors \cite{palacios} and in the studies of condensates 
in rotating anisotropic traps \cite{oktel}.
\begin{figure}
\begin{center}
\epsfig{file=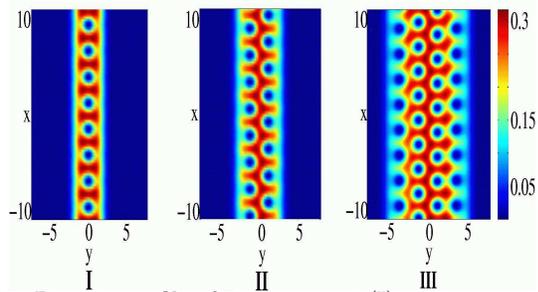,
height=4cm,width=7cm}
\end{center}
\hskip -2cm
\vspace*{-1.2cm}
\caption{Density profile of 2-component (I), 3-component (II), and 
5-component (III) states. (Lengths are in units of $l_0$).
\label{fig:2b}}
\end{figure}

In the presence of many rows of vortices, we may rely on an average 
vortex description in the LLL. Using the relation   
$<\phi_k|y^{2}/l_{0}^{2}|\phi_k> =k^2l_{0}^{2}+1/2$, we write the 
energy functional of the system in the form:
$E[\rho] = \int d^{2}r \left[m \omega_{-}^{2} y^{2} \rho 
+ g \rho^{2}\right]/2$.
Minimizing the energy we obtain the coarse grained density
$\rho(y) = \frac{1}{g}\left[\mu -\frac{1}{2}m \omega_{-}^2 
y^2\right]$, which smooths out density modulations introduced by vortex 
rows. The size of the vortex core is always $\sim l_0$ and the number of the
vortex rows is $\sim\beta^{1/3}$. The energy per particle is given by:
\begin{equation}  \label{E}
\frac{E}{N}\approx \hbar\tilde{\Omega} 
\left(\frac{\omega_-}{2\Omega}\right)^{2}\frac{3}{5}(3 
\sqrt{2 \pi} 
\beta/4)^{2/3}. 
\end{equation}
The result of Eq.(\ref{E}) deviates by less than $10\%$ from the energy obtained by
the full numerical minimization.

The vortex lattice is expected to melt when the number of vortices $N_v$ approaches 
the number of particles $N$ \cite{cooper,melting}.
In our case this condition leads to    
$m^* g/\hbar^{2} \sim n^2 l_{0}^{2}$.
However, our approach of the weakly interacting gas requires $ m^* 
g/\hbar^2 \ll 1$ and, as we are considering large values of $n l_0$, it
breaks down before the melting transition.
Therefore, the description of this 
transition requires a more elaborate treatment of 
quantum fluctuations.

In conclusion, we have shown that a 2D BEC at extreme rotation 
frequency in an elliptic trap forms a novel quantum fluid
in a narrow channel. The  behavior of this fluid is determined by the parameter
$\beta$ Eq.(\ref{beta}) which increases with the interaction strength and effective mass $m^*$. 
For $\beta < \beta_c=4.9$, the
excitation spectrum of the fluid has a small sound velocity and exhibits a
roton-maxon character. At a critical interaction
strength $\beta_c$, the roton
energy reaches zero and the uniform (in the long direction) ground state
becomes unstable. The system then undergoes a second order quantum
transition to the state with a periodic structure which can be viewed
as rows of vortices. For $\beta>\beta_c$, with increasing the interaction parameter $\beta$, 
more vortex rows are nucleated. Finally the vortices form the Abrikosov lattice 
which can melt due to quantum fluctuations.

We acknowledge discussions with J. Dalibard, A. L. Fetter, 
Tin-Lun Ho, J. Palacios, and J.T.M. Walraven. This work was supported by the Minist\`ere de 
la Recherche (grant ACI Nanoscience 201), 
by the Centre National de la Recherche Scientifique (CNRS), by the 
Nederlandse Stichting voor Fundamenteel Onderzoek der Materie (FOM), and in part
by the National Science Foundation (Grant No. PHY99-07949). 
LPTMS is a mixed research unit of CNRS and Universit\'e Paris Sud.

\end{document}